*Research Article*

# Postmortem Analysis of Decayed Online Social Communities: Cascade Pattern Analysis and Prediction

**Mohammed Abufouda**

*Algorithm Accountability Lab, Computer Science Department, University of Kaiserslautern, Germany*

Correspondence should be addressed to Mohammed Abufouda; abufouda@cs.uni-kl.de

Recently, many online social networks, such as MySpace, Orkut, and Friendster, have faced inactivity decay of their members, which contributed to the collapse of these networks. The reasons, mechanics, and prevention mechanisms of such inactivity decay are not fully understood. In this work, we analyze decayed and alive subwebsites from the Stack Exchange platform. The analysis mainly focuses on the *inactivity cascades* that occur among the members of these communities. We provide measures to understand the decay process and statistical analysis to extract the patterns that accompany the inactivity decay. Additionally, we predict cascade size and cascade virality using machine learning. The results of this work include a statistically significant difference of the decay patterns between the decayed and the alive subwebsites. These patterns are mainly cascade size, cascade virality, cascade duration, and cascade similarity. Additionally, the contributed prediction framework showed satisfactorily prediction results compared to a baseline predictor. Supported by empirical evidence, the main findings of this work are (1) there are significantly different decay patterns in the alive and the decayed subwebsites of the Stack Exchange; (2) the cascade's node degrees contribute more to the decay process than the cascade's virality, which indicates that the expert members of the Stack Exchange subwebsites were mainly responsible for the activity or inactivity of the Stack Exchange subwebsites; (3) the Statistics subwebsite is going through decay dynamics that may lead to it becoming fully-decayed; (4) the decay process is not governed by only one network measure, it is better described using multiple measures; (5) decayed subwebsites were originally less resilient to inactivity decay, unlike the alive subwebsites; and (6) network's structure in the early stages of its evolution dictates the activity/inactivity characteristics of the network.

## 1. Introduction

In recent years, online social networks (OSNs) have proven their aptitude as a new medium for sharing news and knowledge, expressing opinions, finding jobs, and many other things. In the literature, there are many works that focus on the growth dynamics of a network, starting with the seminal works of Barabásei and Albert [1] and Watts and Strogatz [2], which were the basis for the field of network science, via many studies examining the growth dynamics of social networks [3–6] to community membership evolution [7], which provide methods and models for analyzing and understanding growth dynamics in social networks. Nevertheless, the dynamics of members' interactions in social networks is not always growth dynamics; many online social platforms have gone through *decay* dynamics in terms of low activity among their members and/or members leaving or deleting their accounts. Online social platforms such as MySpace and Orkut are now out of service after being very active for years and are examples of decayed online social networks. This phenomenon has not been studied well in the literature; decay causes, mechanics, and prevention of decay are still open questions that need to be answered.

Here, we approach the decay dynamics problem from a network perspective by modeling the members as network nodes and their social interactions as temporal edges. We aim to better understand the patterns that occur during the decay process by investigating what we call *inactivity cascades*, which were extracted from decayed Stack Exchange subwebsites. These inactivity cascades are mainly constructed from the structure of the modeled network, where the network structure has already shown to be crucial in understanding the dynamics of any process that takes place on top of a network such as the structure of the World Wide Web networks [8, 9] and social network analysis [10–13]. Moreover, network structure is correlated in many studies to understanding the dynamics of the processes over networks such as epidemic dynamics [14, 15], knowledge spread [16], and knowledge transfer [17]. The information produced and evolved on the Stack Exchange website as an information exchange platform makes this work also connected to the *information dynamics* area [15, 18, 19], where we are concerned in the decay of the information production process on the Stack Exchange website as a medium of knowledge production and sharing.

Based on that, the contributions of this work are summarized as follows:

(i) Extracting and analyzing inactivity cascades from the decayed and alive subwebsites of Stack Exchange

(ii) Devising measures for understanding the decay process and possible patterns in both decayed and alive subwebsites

(iii) Finding different inactivity patterns in alive and decayed subwebsites

(iv) Finding empirical evidence that an inactivity cascade is not driven by only one network measure

(v) Building a machine learning framework for predicting the size and virality of inactivity cascades

The previous contributions can be seen as two parts: (1) *analysis* of the decay process via cascade modeling and (2) *prediction* of cascades' properties. These two parts are complementary because the analysis without prediction limits our control over these platforms and also predicting the properties of a decay requires a better understanding of the decay process itself so that we can provide a good prediction model.

The remainder of this paper is structured as follows. Section 2 describes the related work and highlights how this work contributes to the literature. Section 3 provides the definitions and the methods used throughout this paper, and Section 4 describes the datasets used and some preliminary analyses of these datasets. A detailed description of the results and the prediction framework are provided in Section 5. In parallel to the results, Section 5 also includes a discussion of the results and conclusions of this work. Section 6 presents the limitations of this work and directions for future research.

## 2. Related Work

This paper is related to studies and works that are concerned with decay or inactivity dynamics in social networks. In this section, we present the related works and show how this work is compared to them.

Due to limitations on existing data about interaction decay, researchers have focused on theoretical work based on random networks. For example, Dorogovtsev and Mendes [20] presented a model for understanding the properties of random networks if edges are removed, signaling that the dynamics of a network is not limited to adding nodes and/or edges. Later, with the rise of many social networks and social platforms, research primarily focused on growth dynamics, with very few works dealing with decay dynamics. Torkjazi et al. [21] studied users' migration from MySpace to Facebook when the latter was getting more attention from users. Their study suggests that OSNs have a life cycle that may end with service decay. Dev et al. [22] studied the reasons behind the failure of what they call *knowledge markets*, such as Stack Exchange. They utilized economic production models in order to understand the dynamics of knowledge generated on these knowledge markets. Wu et al. [23] predicted the activity and inactivity of members of the DBLP coauthorship dataset by modeling the dynamics of the social engagement of the members of DBLP. They also provide insights regarding the characteristics of the members who departed the networks using network measures. Similarly, Fenner et al. [24] contributed a theoretical model for generalizing the *rich-get-richer* model of network evolution, which focuses mainly on growth dynamics, by extending it to link deletion in the Web network. Their model implicitly assumes that dynamics is not limited to growth dynamics but may include link removal. Asur et al. [25] approached the activity of users from trend analysis perspective in Twitter, shedding light on what causes some tweets to be trendy. They also found that the decay dynamics of a trend follows a linear function.

Community activity has also been studied by Kairam et al. [26]; they provide machine learning prediction models to predict community longevity. The authors also provide insights into the factors that contribute to keeping online communities active. In the same vein, Abufouda [27] contributed machine learning prediction models for predicting users who left decayed and alive communities, with a focus on the decay dynamics of online communities. Cannarella and Spechler built an epidemic model for predicting the dynamics of the members of Facebook [28]. The results showed that Facebook would lose 80% of its users between 2015 and 2017 (the same model was used by Facebook researchers and predicted that Princeton University would lose half of its students by 2018, see https://www.facebook.com/notes/mike-develin/debunking-princeton/10151947421191849/). Decay dynamics also raised some computational aspects of the decay dynamics problem. Bhawalkar et al. [29] and Zhang et al. [30] provided a theoretical model and mathematical framework for finding the set of nodes whose deletion generates the smallest $k$-core subgraph of a network, focusing on the computational challenge of the decay. Their works assure that the node removal problem is relevant in social and other networks. Ribeiro [31] studied user activity and inactivity by providing a model that uses the number of daily active users as an indicator of the dynamics in membership-based websites. This author also presented a prediction model for predicting whether a community will continue to grow or not, similar to the work in [26]. Malliaros and Vazirgiannis [32] provide a model for social engagement describing the activity and inactivity of members of social networks based on game theory. Similar to the work in [32], Garcia et al. [33] investigated the decay of the Friendster social network using game theory. As one of the results of their work, Garcia et al. argue that decay has a direction, which starts from nodes with less coreness; this was later refuted by Seki and Nakamura [34], who provide a model that shows that decay starts from nodes with higher coreness. Abufouda and Zweig [35, 36] presented a stochastic model for describing the mechanics of inactivity cascades. The model has optimization guarantees that make controlling the decay computationally viable.

The previous works fall into two categories: (1) works that consider both growth and decay processes as a common behavior of online social networks and (2) works that approach the decay process in *social context* only via models, which were not validated with real inactivity decayed data using temporal snapshots. Although the first category seems to be more realistic, none of the related work in this category provides any thorough analysis of the mechanics of the decay process compared to the rich analysis of growth dynamics. This means there is little insight into the decay process of online social interaction, which would serve to better understand online behavior. As a result, the second category of the related work realized that decay dynamics needs to be considered as a separate process and requires further thorough investigation, particularly after the decline of many online social networks like MySpace and Friendster. However, these works used either synthesized data, which led to contradictory conclusions on the same research question (see the work in [33] and an opposing argument in [34] regarding the decay direction and our attempt to resolve this issue in Section 5), or did not consider the temporal aspect of the problem. This study fills the gap by focusing only on decay dynamics using real temporal data from decayed online social communities. Furthermore, we enhance the analysis using inactivity cascades, which, to the best of our knowledge, have not been covered before. This enables us to better understand the characteristics of real inactivity cascades and, hence, helps us gain more insights into the online behavior of humans.

## 3. Definitions and Methods

*3.1. Networks and Measures.* An undirected graph $G$ is defined as a tuple $(V_G, E_G)$, where $V_G$ is the set of nodes of $G$ and $E_G$ is the set of edges that is defined as $E \subseteq V_G \times V_G$. An edge $e = \{v, u\}$ is defined as a pair of two nodes $u$ and $v$, where $u, v \in V$. Graphs at a specific point of time are denoted as $G_t = (V_t, E_t)$, where $V_t$ and $E_t$ are the set of nodes and edges that are observed at time point $t$, respectively. The set of graphs $\mathbf{G} = \{G_0, G_1, \cdots, G_{k-1}\}$ is a temporal structure of a graph at time points $\{0, 1, \cdots, k-1\}$, where $|\mathbf{G}| = k$. The graph $G_0$ is called the *initial network*, where $V_G \subseteq V_{G_0}$, $\forall G \in \{G_1, G_2, \cdots, G_{k-1}\}$.

A *tree* is a connected graph with no cycles. An *inactivity cascade tree* $\mathcal{I}$, a *cascade* for short, is a rooted tree where each directed edge $e = (u, v)$ contains two nodes such that *the last observed time* points of nodes $v$ and $u$ were $\tau(v) = t'$ and $\tau(u) = t''$, respectively, such that $t' > t''$ and $e \in E_{G_0}$. Algorithm 1 describes the steps we followed to extract such cascades. That is, node $u$ became inactive before its neighbor, node $v$. The root of a cascade $\mathcal{I}$ is called a *cascade initiator*, which is any node that becomes inactive while all of its neighbors are active. If no such node exists, we arbitrarily select one of the earliest nodes that became inactive. The number of nodes in a cascade is called cascade *size*.

The edge formation period for an edge $e = (u, v)$, where $e \in E_\mathcal{I}$, is defined as $\tau(v) - \tau(u)$. Based on that, we measure the normalized *cascade duration*, which is defined as

$$\text{CD}_\mathcal{I} = \frac{1}{k \cdot |E_\mathcal{I}|} \sum_{e=(u,v)} \tau(v) - \tau(u). \quad (1)$$

For the set of graphs $\mathbf{G}$, a set of inactivity cascade trees $\mathbf{I}$ is extracted. The *virality* of a cascade $\mathcal{I}$ measures how far the effect of the initiator of a cascade goes [37]. The measure is defined as (this measure was originally proposed as *Wiener index* [38])

$$v(\mathcal{I}) = \frac{1}{n(n-1)} \sum_{v,u \in V_\mathcal{I}} d(u, v), \quad (2)$$

where $d(u, v)$ is the length of the shortest path between the nodes $u$ and $v$ and $n$ is the number of nodes in a cascade. We propose a Jaccard-like similarity measure of two cascades. To have more structural similarity, we consider the structural properties of a cascade by considering the neighborhood of nodes in cascades such that if there is a node shared between two cascades with also many shared neighbors, then the two cascades are assumed to be more similar. Thus, we define

$$\text{sim}(\mathcal{I}_1, \mathcal{I}_2) = \frac{1}{|V_{\mathcal{I}_1} \cap V_{\mathcal{I}_2}|} \sum_{z \in V_{\mathcal{I}_1} \cap V_{\mathcal{I}_2}} \frac{|N(Z_{\mathcal{I}_1}) \cap N(Z_{\mathcal{I}_2})|}{|N(Z_{\mathcal{I}_1}) \cup N(Z_{\mathcal{I}_2})|}. \quad (3)$$

In addition, we used the features in Table 1 for building a supervised machine learning model for predicting cascade's properties.

*3.2. Statistical Divergence*

*3.2.1. Cumulative Distribution Function.* The *cumulative distribution function* (CDF) for a discrete random variable $X$ is defined as $F_X(x) = P(X \leq x) = \sum_{t \leq x} f(t)$. If $X$ is continuous, then the CDF is defined as $F_X(x) = \int_{-\infty}^{x} f_X(t) dt$. Similarly, the complementary CDF is defined as $\bar{F}_X(x) = P(X > x) = \sum_{t > x} f(t)$.

*3.2.2. Kolmogorov-Smirnov Test.* The *Kolmogorov-Smirnov test* (KS-test) is a statistical test that tells whether two different samples were drawn from the same distribution or not. The test is used to compare two patterns in order to know

```
    Input: G_0, G_1, ..., G_{k-1} //The set of temporal networks
    Init: I = ∅, S = {L_1, L_2, ..., L_{k-1}}
    //L_i ∈ S is defined as: {v | τ(v) = i, ∀v ∈ V_{G_0}}. L_1 contains the initiators of the cascades
 1  foreach v ∈ L_1 do
 2      𝒥 = (V_𝒥 = {v}, E_𝒥 = ∅)//Start a new cascade 𝒥
 3      //Check if the initiator v is connected to another initiator q
 4      foreach q ∈ L_1 do
 5          if   e = {v, q} ∈ E_{G_0} then
 6              V_𝒥 = V_𝒥 ∪ {q}//Add the node q to the cascade's nodes
 7              E_𝒥 = E_𝒥 ∪ {e = (v, q)}// Add a directed edge (v,q) to the cascade
 8          end
 9      end
        //Check if the initiator v is connected to any non-initiator node
10      foreach L ∈ S \ {L_1} do
11        foreach u ∈ L do
12          if e = {v, u} ∈ E_{G_0} then
13              V_𝒥 = V_𝒥 ∪ {u}
14              E_𝒥 = E_𝒥 ∪ {(v, u)}
15          end
            //Check if u is connected to any other nodes in the cacasde 𝒥
16          else
17              foreach w ∈ V_𝒥 do
18                  if e = {u, w} ∈ E_{G_0} then
19                      E_𝒥 = E_𝒥 ∪ {e = (u, w)}
                        //The break prevents triangles from being formed in 𝒥
20                      break
21                  end
22              end
23          end
24        end
25      end
        //Add the extracted cascade 𝒥 to the set of all cascades I
26      I = I ∪ {𝒥}
27  end
    Output: I
```

ALGORITHM 1: The steps for extracting inactivity cascades, **I**, from the set of temporal networks networks **G**.

if they are the same or statistically different. Informally, it is the maximum absolute distance between the two CDFs of the two samples. More formally, for two CDFs, $F_1$ and $F_2$, the KS-test statistics $D$ is defined as $D_{KS} = \sup_{-\infty<x<\infty}|F_1(x) - F_2(x)|$, where $\sup_{-\infty<x<\infty}$ is the supremum of a set.

*3.2.3. Patterns' Entropic Similarity.* Shannon entropy [39] quantifies the information in a discrete random variable $x \sim p(x)$ as follows: $H(P) = -\sum_{i=1}^{n} p(x_i) \cdot \log p(x_i)$. Given two probability distributions $P$ and $Q$, the *Kullback-Leibler divergence* [40] ($D_{KL}$) is a measure that finds how similar these two distributions are and it is defined as $D_{KL}(P\|Q) = \sum_{i=1}^{n} p(x_i) \cdot \log p(x_i)/q(x_i)$. The *Jensen-Shannon divergence* is then defined as $D_{JS}(P, Q) = 1/2[D_{KL}(P\|R) + D_{KL}(Q\|R)]$, where $R = 1/2(P + Q)$, which is a symmetric distance variation of the $D_{KL}$.

## 4. Dataset

Stack Exchange (https://StackExchange.com/) is a network of questions and answer websites that contain subwebsites for specific topics, such as computer science, German language, or workplace, to name just a few. Before being available to the public permanently, each of these websites must go through a beta version, becoming permanent for the public if it sustains a certain level of activity. If the subwebsite does not meet the activity requirement, it is shut down. Some of these subwebsites go back and forth between being beta and closed. As a result, all of the users' accounts and their interactions are saved. This information is the dataset used for this work. We parsed, structured, and analyzed a set of closed (decayed) subwebsites as an example of communities that underwent decay dynamics [27, 36] and alive subwebsites, respectively. The decayed subwebsites we considered in this work are *Business Startups* and *Economics*. In addition to that, we also have data for alive websites, such as *Statistics*, *Latex*, and *Music*. We used both types in order to make a comparison, if possible, between the patterns and cascades found in the alive and the decayed communities. One advantage of this dataset is that it contains all the temporal information needed to construct temporal social networks based on the interactions among the users. So, we constructed networks where the nodes are the members of these networks and the edges are the interactions among them, including replying to a question, upvoting, or downvoting. Table 2 shows a summary of the datasets used, the monitoring period for the interactions, the number of networks constructed, information about the first and the last networks, and the number of extracted cascades. The monitoring period for the datasets differed according to their active periods; e.g., for the decayed subwebsites (the first three rows in Table 2), the last monitoring day was the last day these subwebsites were active. Conversely, the last three subwebsites are still alive, so the last monitoring day was the same. Note that the set of nodes $V_{G_0}$ refers to the core nodes used for this study, which means other nodes emerging in-between were ignored. The core nodes were members with a reputation score of at least 500; we tried smaller values, e.g., 100, 200, 300, and 400, for the reputation score, and the resulting temporal networks were too sparse with too many disconnected components which hinders any subsequent analysis. The reason for that in the context of the Stack Exchange websites is that there are many users who come only for one question or make only one comment and then do not appear again on the platform. We consider those users as outliers to the platform's core activity, e.g., information production, and thus, the chosen value, i.e., reputation score ≥500, is justified from the lower bound side. We did not select larger values for two reasons: (1) there are few users in some communities who have reputation score larger than 500 and (2) selecting larger reputation score excludes members with less activity and thus the core nodes become significantly few nodes. Both cases render the constructed networks useless for any analysis. Thus, the chosen value is justified from the upper bound side.

Table 1: Definitions of the network-based measures used in this work.

| Measure | Description |
| --- | --- |
| $D(v)$ | The *degree* of a node $v$, $D(v) = |\Gamma(v)|$, is the cardinality of the set of neighbors $\Gamma(v)$. |
| $B(v)$ | The *betweenness* of a node $v$ is defined as $B(v) = \sum_{s \in V(G)} \sum_{t \in V(G)} \sigma_{st}(v)/\sigma_{st}$, where $\sigma_{st}(v)$ is the number of the shortest paths between nodes $s$ and $t$ that include the node $v$ and $\sigma_{st}$ is the number of all the shortest paths between nodes $s$ and $t$. |
| $\mathscr{C}(v)$ | The *closeness* of a node $v$ is defined as $\mathscr{C}(v) = (\sum_{w \in V(G)} d(v,w))^{-1}$, where $d(v,w)$ is the distance between nodes $v$ and $w$. |
| $\text{Core}(v)$ | A *k-core* subgraph of a graph $G$ is the maximal subgraph such that each node has a degree at least $k$. The *coreness* of a node $\text{Core}(v) = k$ if the node $v$ is in the $k$-core subgraph and not in the $k+1$-core subgraph. |
| $E(v)$ | The *eccentricity* of a node $v$, $E(v)$, is the maximum distance between node $v$ and node $u$. |
| $\mathscr{MC}(v)$ | A *minimum cut* of two nodes $u, v$, $\text{MinCut}(u,v)$ is the minimum number of edges that are required to be removed in order to separate the two nodes. The averaged minimum cut of a node $v$ is defined as $\mathscr{MC}(v) = 1/n \sum_{u \in E, u \neq v} \text{MinCut}(u,v)$, where $n$ is the number of nodes in a graph. |
| $\text{Evec}(v)$ | The *eigenvector centrality* of a node which is defined as $\text{Evec}(x_i) = 1/\lambda \sum_{j \in V_G} a_{ij} x_j$, where $\lambda$ is a constant and $a_{ij}$ is a location defined by $i, j$ in the adjacency matrix. The measure can be written in matrix form as $\lambda \mathbf{x} = \mathbf{A} \cdot \mathbf{x}$. |
| $B(e)$ | *Edge betweenness* measures the number of times an edge $e$ appears in the shortest path between any two nodes in a graph. It is defined as $B'(e) = \sum_{v,u \in V_G} \sigma_{uv}(e)/\sigma_{u,v}$. The *incident edge betweenness* of a node is defined as the average edge betweenness for all edges incident to a node $v$. |
| $D(\Gamma(v))$ | The average degree of the neighbors of a node $v$. |

Table 2: Description of the datasets used and the $k$ constructed networks over the given period. The initial network is $G_0$, and the last network is $G_{k-1}$. The number of the extracted cascades is $|\mathbf{I}|$.

| Dataset | Period | $k$ | $|V_{G_0}|$ | $|E_{G_0}|$ | $|V_{G_{k-1}}|$ | $|E_{G_{k-1}}|$ | $|\mathbf{I}|$ |
| --- | --- | --- | --- | --- | --- | --- | --- |
| Startups | 10.2009–09.2013 | 32 | 702 | 9080 | 2 | 1 | 309 |
| Literature | 08.2011–05.2012 | 12 | 118 | 434 | 4 | 3 | 4 |
| Economics | 10.2011–03.2012 | 10 | 33 | 67 | 3 | 2 | 17 |
| Latex | 07.2010–12.2015 | 33 | 498 | 4823 | 53 | 87 | 169 |
| Statistics | 07.2010–12.2015 | 32 | 419 | 4795 | 36 | 37 | 141 |
| Music | 04.2011–12.2015 | 38 | 293 | 1303 | 2 | 1 | 48 |

From the table, it is clear that the alive subwebsites Latex and Statistics, which are considered very active, succeeded in keeping nearly 10% of the core nodes in the last network, whereas this percentage is almost *zero* in the other subwebsites. We found that those 10% of the members were users with very high overall reputation score. For instance, user number 5001 (https://tex.stackexchange.com/users/5001/mico) was active in all of the networks used overtime for the Latex subwebsite, and he/she is in the top 0.09% among the whole users of Stack Exchange and have reputation score 236 thousands. The same behavior was found on the Statistics subwebsite for user 805 (https://stats.stackexchange.com/users/805/glen-b) who is in the top 0.02% among the Stack Exchange users and have reputation score 191 thousands. We noticed that these two users were active mainly on the corresponding subwebsite, Latex and Statistics, respectively. For the Music subwebsites, the situation is different. The number of retained members from the core nodes was only two users, which is very similar to the decayed subwebsites. Moreover, those two users were mainly active on other subwebsites; for example, user 932 (https://music.stackexchange.com/users/932/leftaroundabout) was found in all of the networks of the Music dataset, but his main activity was on the Stack Overflow subwebsites. For the decayed websites, it was hard to get information about the retained users from the core users because no user information was available.

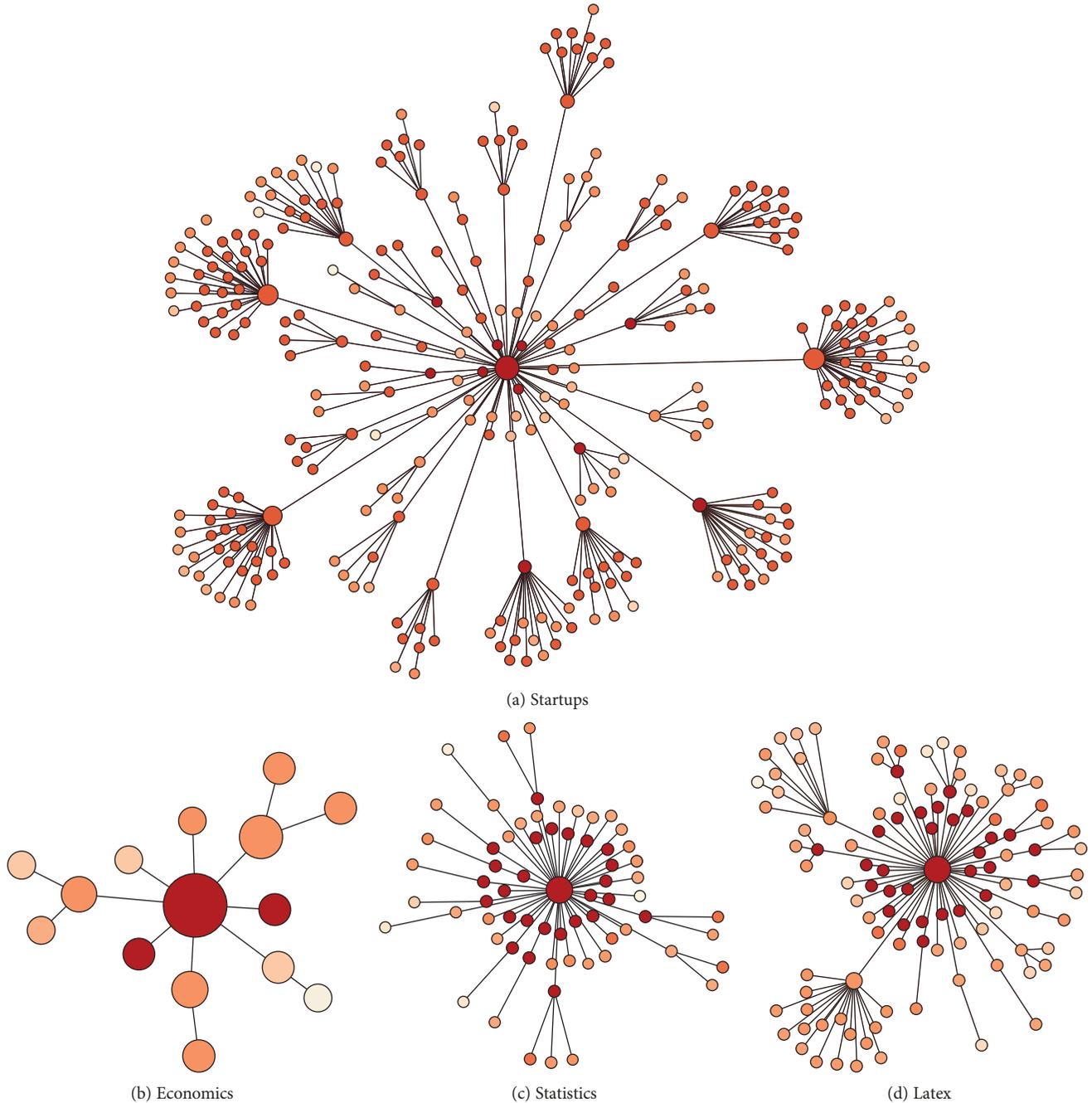

(a) Startups

(b) Economics

(c) Statistics

(d) Latex

Figure 1: The largest cascades extracted from the datasets. The color of the node is inversely proportional to the time at which the node became inactive (i.e., the darker the node, the earlier it became inactive), and the size of a node is directly proportional to its degree in the cascade.

## 5. Results and Discussions

*5.1. Analysis and Modeling Results.* Here, we start presenting the results of the analysis by providing information about the largest cascades extracted from the datasets. Figure 1 shows the largest cascades of the subwebsites Startups, Economics, Statistics, and Latex. We observe that the cascades of the decayed communities, such as Startups and Economics, contain a larger fraction of nodes from the initial network $G_0$. The fraction of the nodes in the largest cascades, considering the initial network, is 0.44, 0.45, 0.15, 0.21, and 0.09 for the subwebsites Startups, Economics, Statistics, Latex, and Music, respectively.

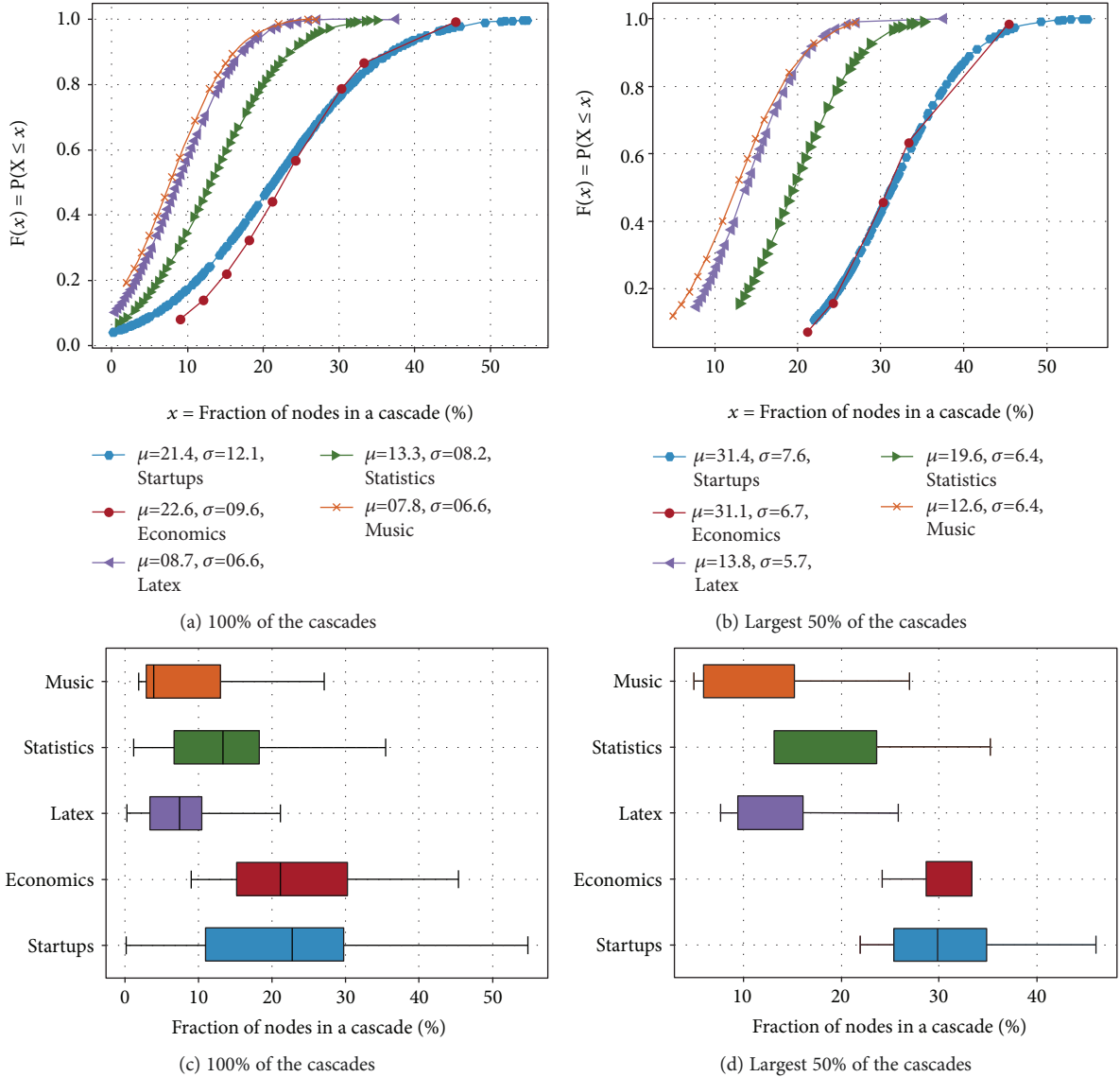

Figure 2: The figure shows the fraction of nodes (of the initial network) in the observed cascades as CDF (a and b) and as box plots (c and d).

The figure also shows that for the decayed subwebsites, the color of the nodes is very close to each other, which suggests that the duration of the decayed subwebsites was short compared to the duration of the alive subwebsites, because the colors of the nodes in the alive subwebsites are clearly lighter at the nodes close to the leaves. This will be statistically supported in the following section.

*5.1.1. Cascade Size.* The size of a cascade is the number of nodes it contains. Figure 2 shows the results obtained for different subwebsites. We can observe in the figure that all datasets contain cascades that have at least 38% of the nodes from the nodes of the initial network. This percentage is even higher in decayed communities (Startups and Economics) and reaches 55% on the Startups subwebsite. Figure 2 also shows that the cascade size patterns appear visually different. The difference is even clearer in Figures 2(b) and 2(d), where the cascades in the decayed communities contain a lot more nodes. To get statistical significance concerning this phenomenon, we used the KS-test described in Section 3.2. We found that there is statistical significance between the decayed and the alive subwebsites. We found that the probability distributions of the cascade size are the same (e.g., seems to be drawn from the same distribution) in the alive websites ($p \approx 0.12$), are the same for the decayed subwebsites ($p \approx 0.7$), and are different when testing an alive website and a decayed website ($p \ll 10^{-6}$). The only exception to this occurred when testing the statistical significance between the Statistics and the Latex subwebsites; although both are still alive, the cascade sizes were statistically different ($p \ll 10^{-6}$).

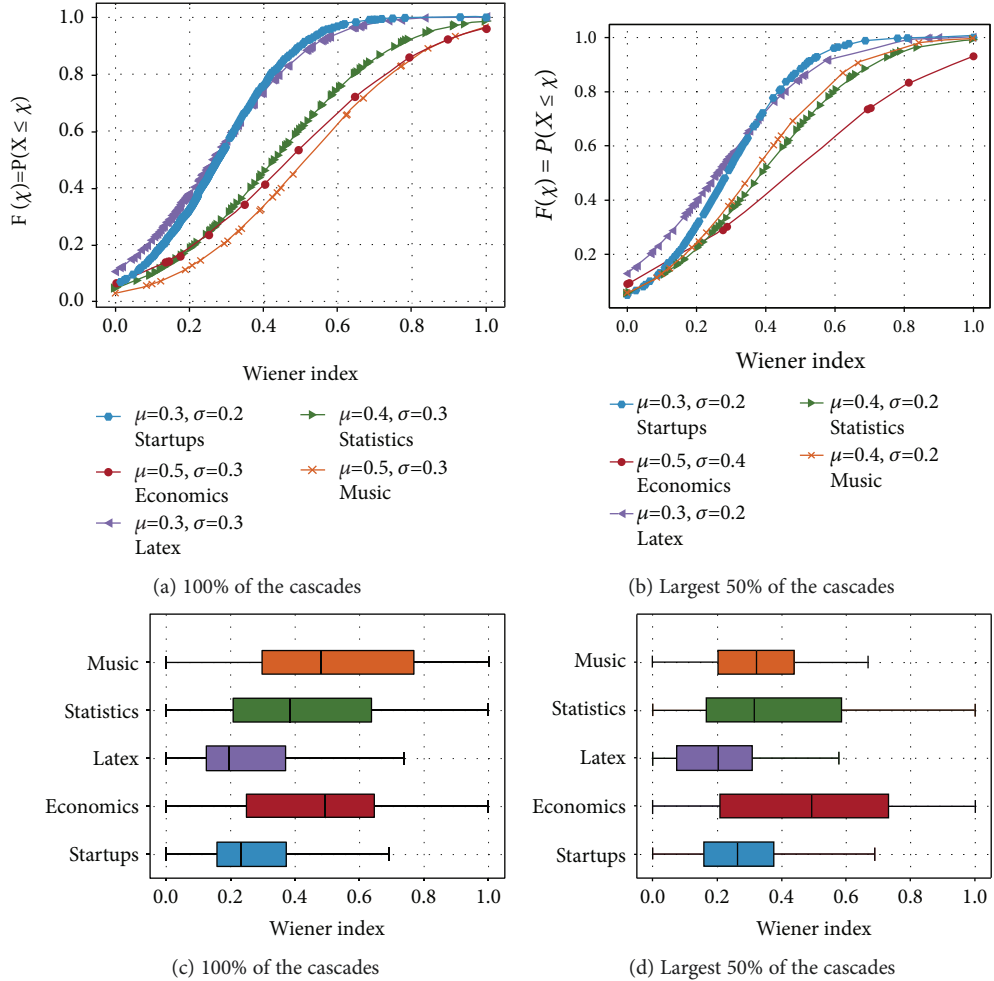

Figure 3: The figure shows the Wiener index of the observed cascades as CDF (a and b) and as box plots (c and d).

*Discussion Point 1.* Different inactivity cascade patterns exist in alive and decayed subwebsites.

The size of the cascades extracted from different subwebsites shows that inactivity dynamics is common in both alive and decayed subwebsites of the Stack Exchange. However, the size of the cascades in the decayed ones was significantly larger than the size of the inactivity cascades found in the alive subwebsites. Based on Figure 2, the smallest cascade in the largest 50% of the cascades contains more than 20% of the nodes from the initial network of the decayed subwebsites, compared to nearly 10% for the alive ones. Our interpretation of this is that there are members of the alive subwebsites who are maintaining the aliveness of these communities and continuously provide content (in terms of, for example, answers to the questions), which keeps the platform active. This can be clearly seen in Table 2, where in the alive subwebsites, the number of nodes found in the last observed network is very much higher than that of the nodes found in the decayed subwebsites. It seems that those members are experts whose existence is vital for sustaining these communities. Investigating the profiles of some of those members (see Section 4) supports our interpretation.

*5.1.2. Cascade Virality.* Figure 3 shows the Wiener index of the cascades extracted from different subwebsites as a measure of virality. As the size of the networks and the size of the cascades differ across the subwebsites, it was necessary to normalize the Wiener index to enable a meaningful comparison of the distributions. To that end, we used a sigmoid function for normalization (other normalization methods like tanh function and min–max normalization provided almost identical results). Generally, the patterns of virality across different subwebsites are statistically the same ($p > 0.1$), except for the Economics subwebsite, where the virality patterns are statistically different with $p \ll 3 \times 10^{-5}$. This special behavior of the Economics dataset is ascribed to it being a small dataset with only 17 cascades. Surprisingly, the figure shows that the decayed subwebsite Startups shows fewer viral cascades, with a mean of 0.27. This suggests that there should be another feature affecting the decay of the decayed subwebsites. In the following section, we will discuss this in more detail.

*5.1.3. Maximum Degree of Cascade.* Another pattern that we looked into is the maximum degree in a cascade. Figure 4 shows the normalized maximum degree in a cascade for different subwebsites. The visualization suggests that the decayed subwebsites Startups and Economics contain cascades of nodes with larger degrees than the alive subwebsites. The statistical analysis shows that the decayed subwebsites have a very similar distribution of the maximum degree in a cascade with $p > 0.13$. The decayed and the alive subwebsites are statistically different with $p \ll 10^{-8}$. Once again, the Statistics subwebsite shows a different pattern. It is neither similar to any of the decayed subwebsites nor to any of the alive subwebsites, with $p \ll 10^{-8}$.

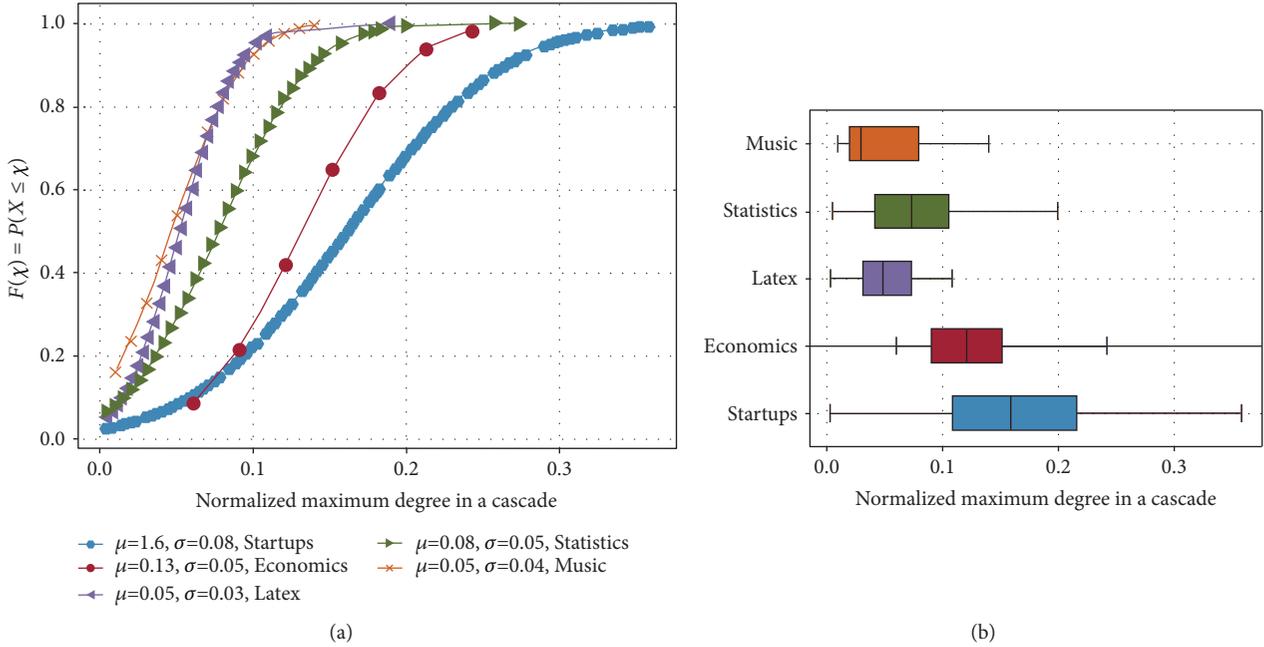

Figure 4: The figure shows the normalized maximum degree of a node in the observed cascades as CDF (a) and as box plots (b).

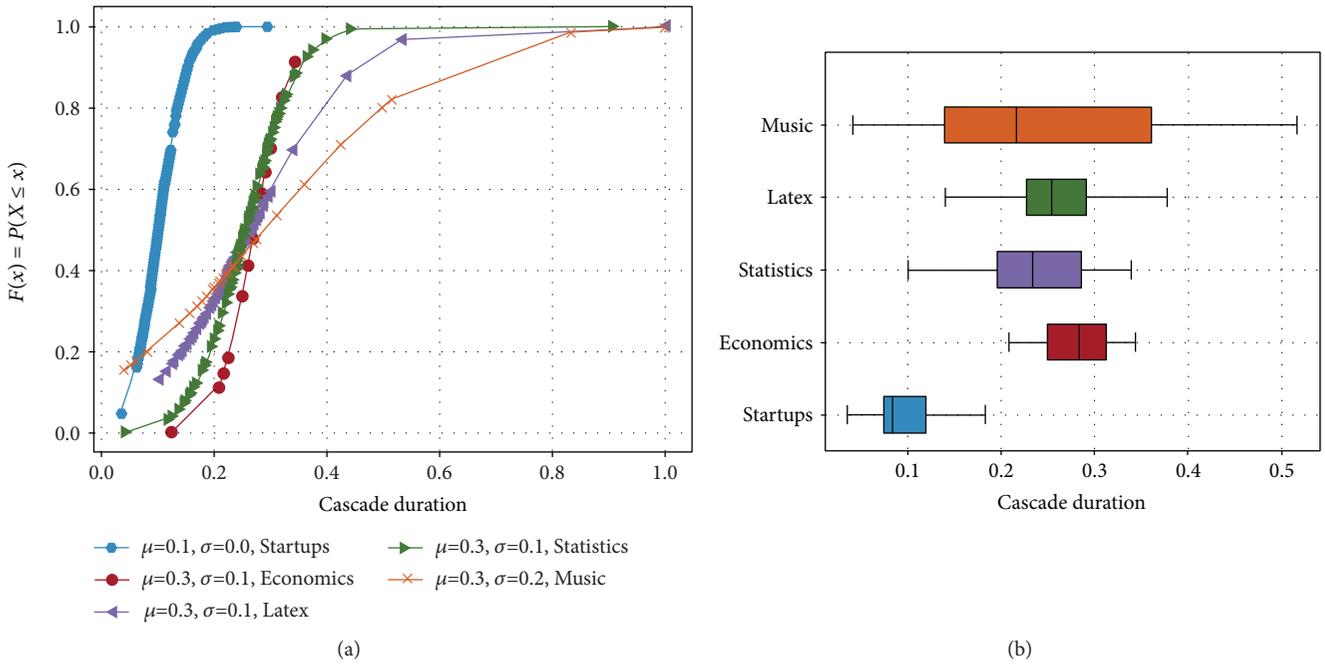

Figure 5: The figure shows the cascade duration as CDF (a) and as box plots (b). Cascade duration is normalized based on the number of networks available for each subwebsite.

*Discussion Point 2.* Inactivity decay is ascribed to a cascade's node degrees, not to its virality.

Unexpectedly, the decayed subwebsites we examined had fewer viral cascades than the alive subwebsites. This led us to investigate the microproperties of the cascades rather than relying only on the macroproperties. We found that the cascades in the decayed subwebsites are less viral, but their nodes have larger degrees compared to those in the alive subwebsites. Additionally, we discovered that cascade initiators in decayed subwebsites have larger degrees in the cascade trees than noninitiators. This indicates that the expert members (who have larger degrees due to their activity and contribution) started the inactivity process, followed by non-expert members. Having said that, one possible reason for the closure of the decayed subwebsites is the lack of activity from those members who should have sustained the community and kept it going until it reached the public version. On the other hand, the more viral cascades in the alive subwebsites, which also have a smaller number of nodes and contain nodes with smaller degrees than the decayed subwebsites, indicate that the effect of inactivity is limited. The reason for this is that the size of the cascades in the alive subwebsites is small, with initiators having smaller degrees, compared to decayed subwebsites. We conclude that expert members acted as obstruction points in the cascade trees, stopping the effect of inactivity cascades from being very disruptive.

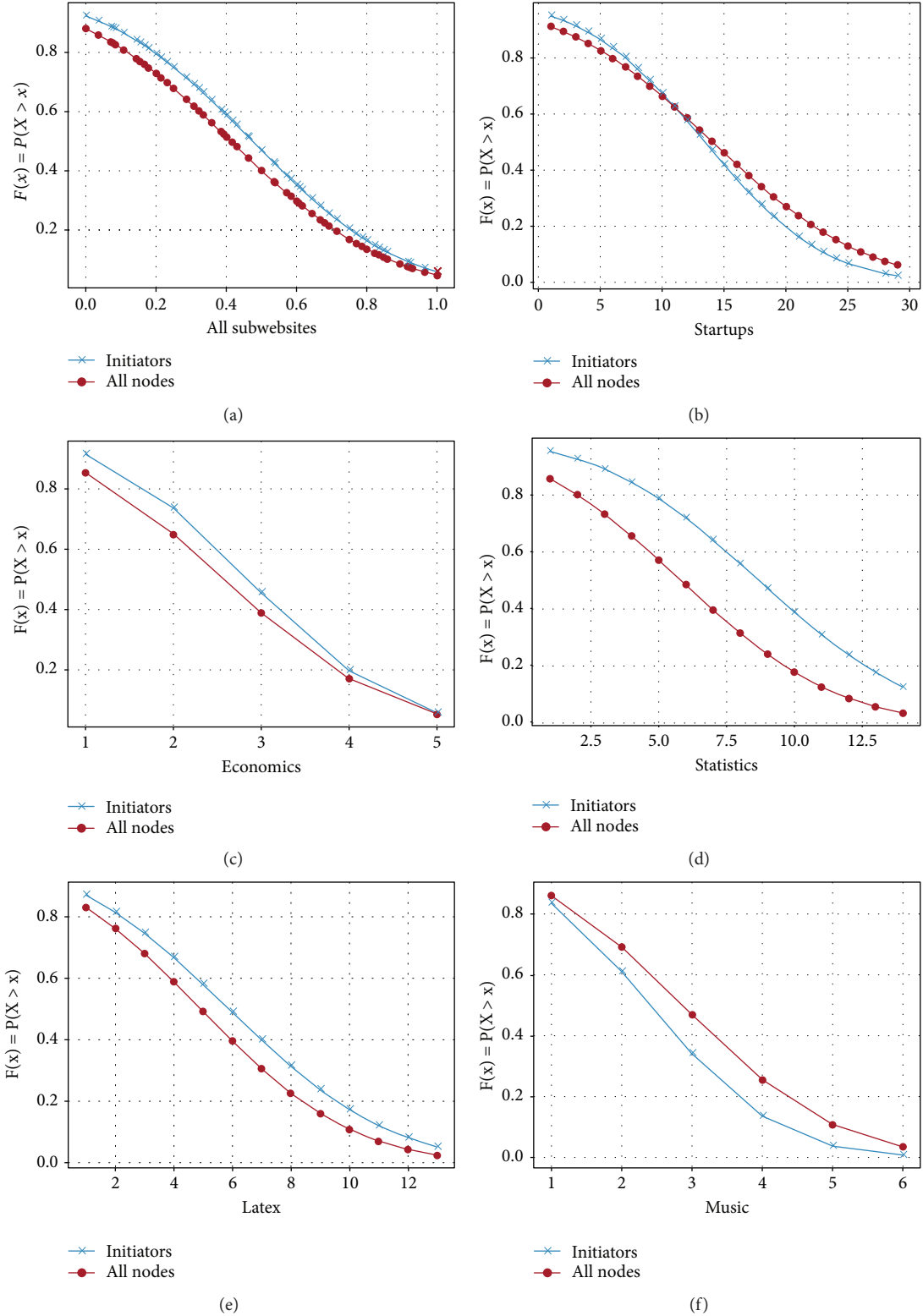

Figure 6: The figure shows the CCDF of the probability distribution of the coreness of all nodes in the network $G_0$ compared to the coreness of the initiators for all subwebsites combined (a). In addition, (b) to (f) show the CCDF for each subwebsite alone. Note that the $x$-axis scale in (a) is scaled to allow for proper comparison, as network sizes are different across different subwebsites.

*5.1.4. Cascade Duration.* Here, we provide the results for the analysis of cascade duration defined earlier in Section 3.1, (1). Figure 5 shows the cascade duration of different subwebsites. The normalized $x$-axis reflects how long the cascade takes to be completed, i.e., until the last day of the observed time. The figure shows that the cascades in the decayed subwebsite Startups took noticeably less time to be completed, i.e., it had faster cascades. This is also clearly visible in Figure 5(a). The statistical analysis of cascade duration showed that every subwebsite has its own characteristics, with no common pattern identified ($p < 5^{-10}$).

*Discussion Point 3.* Which subwebsite is going to decay next?

Although the Statistics subwebsite is alive and falls into the category of alive subwebsites based on the results described in Sections 5.1.1, 5.1.3, and 5.1.4, we discovered that the Statistics subwebsite inactivity patterns are closer to the patterns found in the decayed subwebsites than to those of the other alive subwebsites. Using the $D_{JS}$ described in Section 3.2, we found, strangely, that the Statistics subwebsite is closer to the decayed subwebsites in terms of cascade size, virality, maximum degree in a cascade, and cascade duration. We investigated this behavior and found that the Statistics subwebsite is the least active subwebsite among all Stack Exchange subwebsites with the fewest answered questions; that is, only 61% of the questions were answered (https://stackexchange.com/sites), whereas on other subwebsites, the answer rate is much higher, for example, reaching 93% and 97% on the Latex and Music subwebsites, respectively. This odd behavior, which was caught by our result, supports the effectiveness of the method we used. We think that the Statistics subwebsite may fall into a decay process if its activity level remains as low as it is.

*5.1.5. Cascade Coreness.* Here, we examine the coreness of the nodes in a cascade as a microscopic property of a cascade. We start by examining the coreness of an initiator. Figure 6(a) shows a comparison between the coreness of all noninitiator nodes in network $G_0$ and the coreness of the initiators from all subwebsites as CCDF. The figure shows that the probability of having a coreness, say $x$ in the initiators, is larger than what is found for all nodes. This suggests that the coreness of the initiators is larger than that of the other nodes in the initial network $G_0$. This was also statistically confirmed with $p \ll 10^{-6}$. However, further examination provided different insights and patterns. We performed the same analysis for each of the subwebsites. For example, in Figure 6(b), there was no clearly different pattern for the subwebsite Startups, where the initiators have higher coreness for the coreness values [22, 27] but less coreness for the coreness values [20, 32]. For the other subwebsites in Figures 6(c), 6(e), and 6(d), the initiators have a clear pattern. They have more coreness than the other nodes in the corresponding $G_0$. An opposite pattern was found in the subwebsite Music (cf. Figure 6(f)).

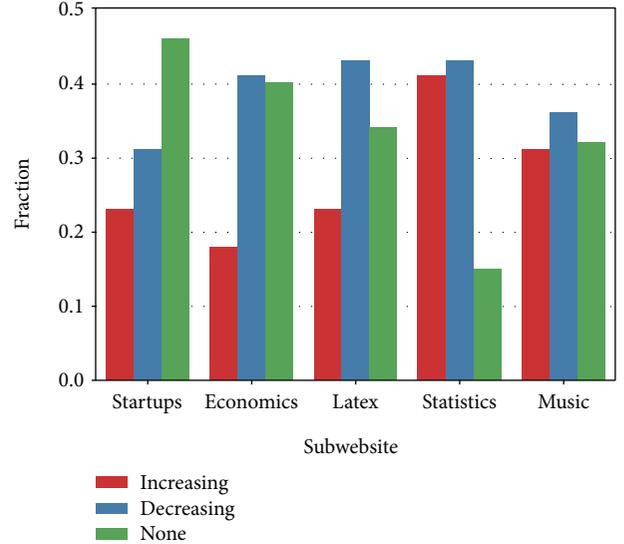

FIGURE 7: The coreness monotonicity of all cascade paths extracted from all cascade trees originating from the cascade initiators. On the $x$-axis, the different subwebsites are shown and on the $y$-axis, the fraction of paths is monotonically increasing, monotonically decreasing, or none monotone.

The previous analysis only refers to the initiators. To understand the coreness in the temporal context, we define the following: a *cascade path* $P$ is a connected directed subgraph of a cascade $\mathcal{I}$, where the maximum degree for all nodes of $P$ is 2, with no cycles. The *coreness monotonicity* of a cascade path $P$ is said to be *increasing* if $\text{core}(v) \geq \text{core}(u)$, *decreasing* if $\text{core}(v) \leq \text{core}(u)$, and *nonmonotone* otherwise, $\forall e = (u, v) \in E_P$. If the nodes in a cascade path have the same coreness, then we consider it nonmonotone. All coreness values are calculated in the initial network $G_0$. Based on that, we extracted cascade paths from all cascade trees where the first node in a path is the initiator of this cascade tree. Then, we examined the coreness monotonicity of these paths. The results are shown in Figure 7 and indicate that the coreness of the cascade paths is clearly different across different subwebsites. Moreover, the fraction of monotonically increasing and monotonically decreasing paths was nearly identical in some cases (see, for example, the Statistics and Music subwebsites). Also, in the case of decayed subwebsites (see the Startups subwebsite), the fraction of nonmonotone paths was larger than for any of the other two types.

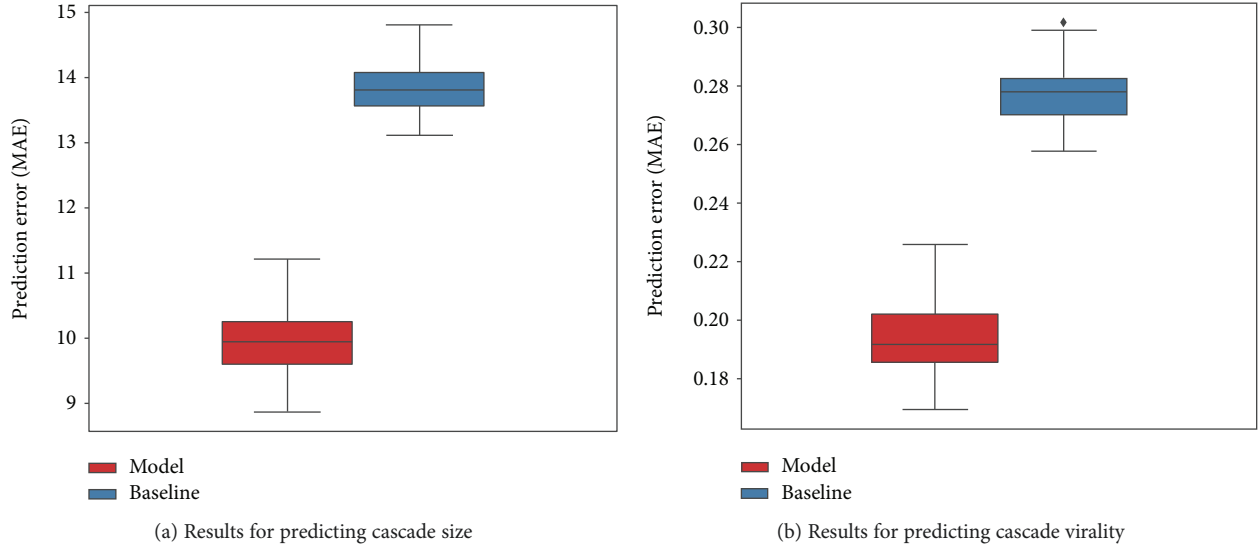

(a) Results for predicting cascade size

(b) Results for predicting cascade virality

Figure 8: The figure shows the prediction performance results for 100 runs for the prediction of cascade size (a) and cascade virality (b). The figure compares the results of the *GBR* prediction algorithm and the results obtained from a baseline predictor.

*Discussion Point 4.* Coreness (and generally speaking, any single measure) alone does not control inactivity cascades.

In their work, Garcia et al. [33] posed the question of whether the decay starts from the *interiors* (nodes with high coreness) or from *exteriors* (nodes with low coreness). In their work, they argued that the decay of the Friendster social network started from exterior nodes. Later, Seki and Nakamura [34] presented a counter-argument, showing that the decay started from the interiors, and provided a model for understanding the decay process. Here, we argue that the answer to the question "Does the decay start from the interior or the exterior nodes?" is neither. The results of this work show no uniform pattern across different subwebsites that correlates to the direction and the coreness of the decay (cf. Figure 7). Furthermore, we argue that the question contains an implicit unsupported assumption, namely, that coreness only controls the decay. We strongly believe that coreness alone cannot be used to understand the direction of decay dynamics if the direction really matters. In Section 5.1.5, we provided a formal framework defining the direction of the decay considering the temporal decay so that we can explicitly tell whether coreness alone can be used as an indicator for the direction of the decay. We found that the initiators of cascades contain opposing patters in terms of whether their coreness is higher or smaller than the coreness of noninitiators. Additionally, we analyzed the coreness of the nodes in the cascade paths (coreness monotonicity) and found evidence that coreness is not correlated with the direction of the decay. Moreover, we performed an analysis using different measures, like degree and betweenness. We conclude that it is very hard to describe the decay process using only one measure. This is also clearly visible in the prediction results (cf. Figure 8) where the importance of the features used for predicting cascade size and virality was close. To further support our argument, we predicted cascade size and virality using only one feature. In no case were the results better than when we predicted them using multiple features. We found the results of prediction using only one feature to be very close to the baseline predictor; for example, the MAE was 0.23, 0.23, 0.22, and 0.22 for predicting cascade virality using betweenness, degree, coreness, and min. cut, respectively. To sum up this point, we think that inactivity decay may be caused by network-independent factors, like privacy issues, competence between social network providers, and/or content quality. If any of these factors manifests itself, it renders the network measures unusable for describing inactivity decay.

*5.1.6. Cascade Similarity.* Using the similarity measure defined in (3), we calculated the similarity of each pair of cascades. Figure 9 shows a heat map for the similarity of the cascades for different subwebsites. Figure 9(a) clearly shows less similarity between the cascades of the Startups subwebsite, unlike the other panels in Figure 9. It is also observed that cascades with a smaller number of nodes seem to be more similar than those with a large number of nodes. An exception is the Economics subwebsite, where cascades with larger nodes are more similar than those with fewer nodes.

To get statistical confidence about the comparison, we used the statistics described in Section 3.2. We found that although all of the subwebsites exhibit different similarity patterns ($p \ll 10^{-8}$), the decayed subwebsite Startups has the smallest average similarity with a value of 0.03, compared to 0.21, 0.16, 0.17, and 0.11 for the other subwebsites. This difference can easily be seen in Figure 10.

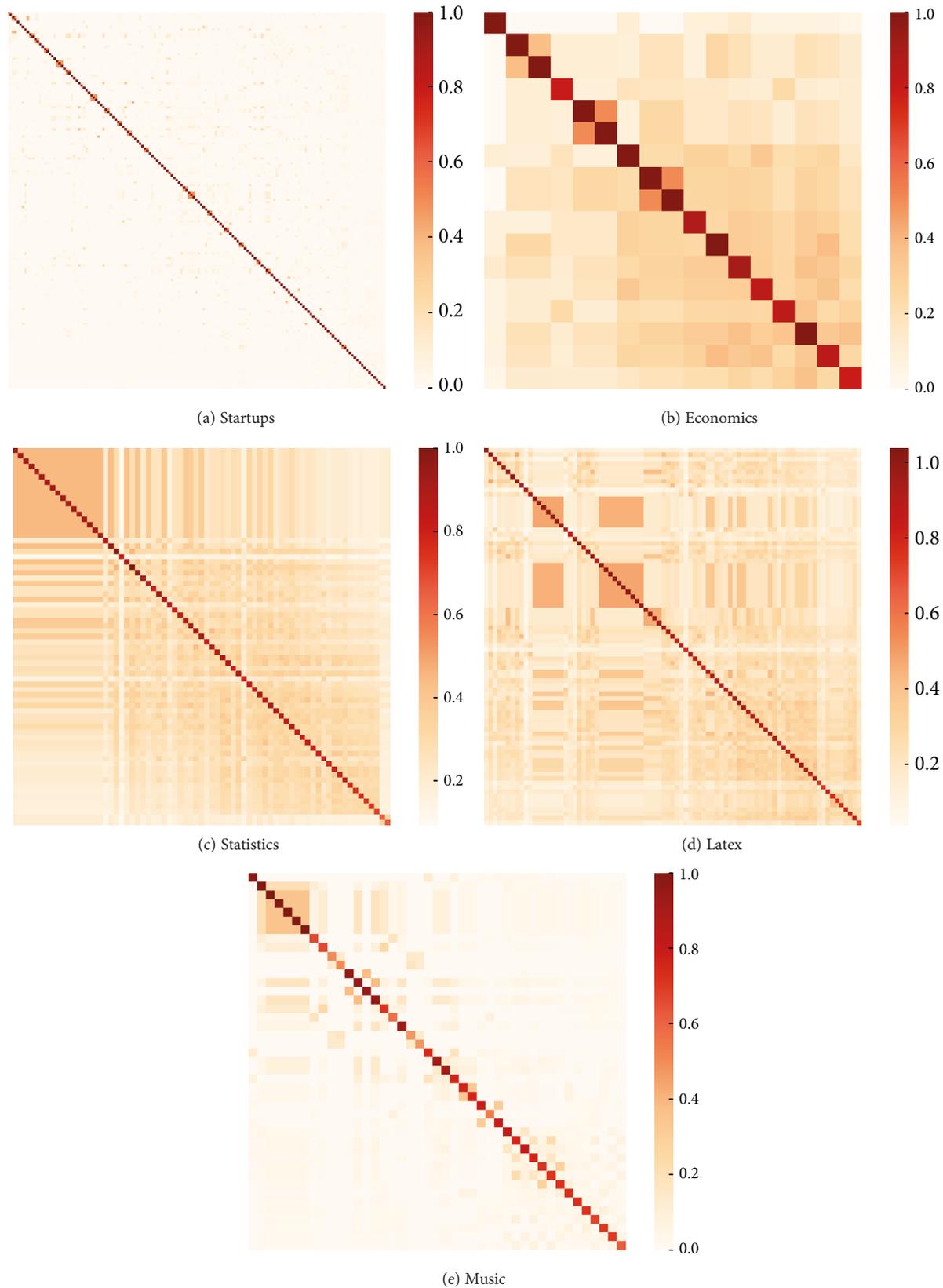

Figure 9: The figure shows the similarity of each pair of cascades for different subwebsites. The cascades were ranked in ascending order based on the number of nodes they have. The darkness of the color is directly proportional to the similarity.

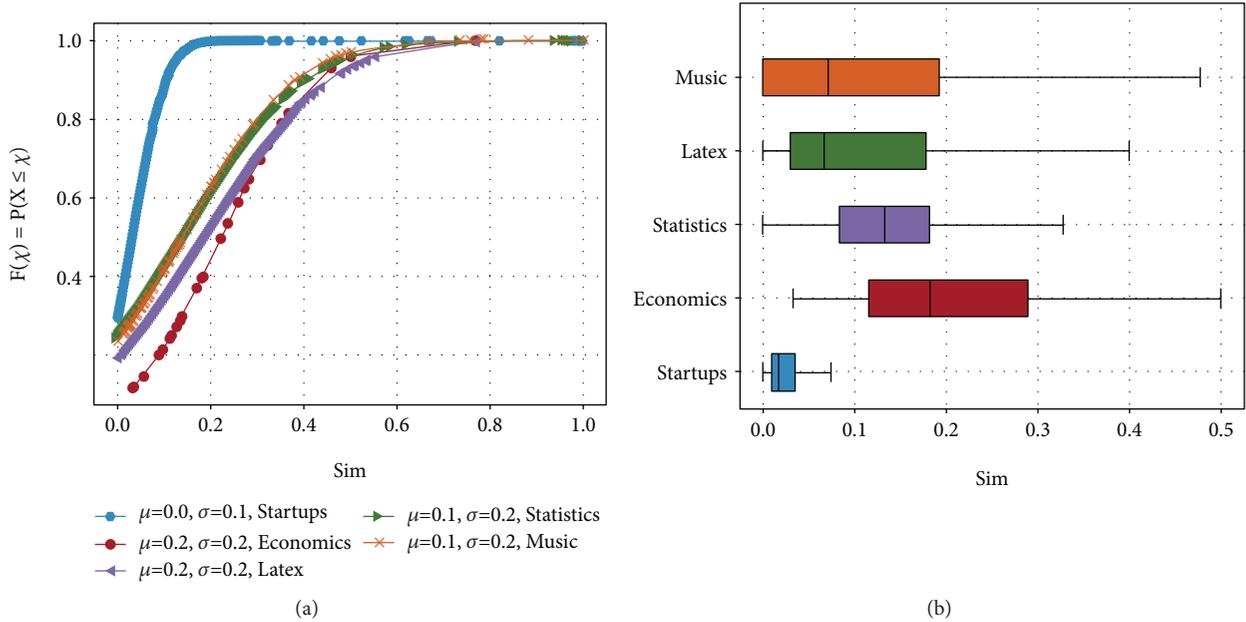

Figure 10: The figure shows the similarity of each pair of cascades as CDF (a) and as box plots (b). The similarity is defined as described in (3).

*Discussion Point 5.* Cascade similarity reflects how resilient a network was while it evolved.

The model we described for the extracted cascades in Section 3.1 allows for cascades with the same nodes and/or edges. This means that we can measure the similarity of two cascades. Basically, if there are many similar cascades in a subwebsite, this means that there are fewer paths on which the inactivity cascade took place than less similar cascades. This means that, for cascades with less similarity, there are many decay propagation paths that are susceptible to inactivity and conversely, for cascades with high similarity, there exist fewer decay propagation paths that are susceptible to inactivity. Thus, cascade similarity can be seen as a measure for the resilience (or vulnerability) of a community for any future model or simulation of inactivity decay. Based on the results described in Section 5.1.6, it is apparent that the decayed subwebsites contain more nodes that are susceptible to inactivity than the alive subwebsites. The similarity of the cascades in the alive subwebsites is high, suggesting a lower number of cascade paths.

*5.2. Prediction Results.* In this section, we provide a prediction framework we designed for predicting some cascade features. We formalize the prediction problem as follows. Given a training set $Z = \{(X_1, y_1), \cdots, (X_n, y_n)\}$, where $X_i = \{x_1, \cdots, x_m\}$ is the set of input features of length $m$, $y_i$ is the target value to be predicted, and $n$ is the number of data points in the training set. The prediction problem is then defined as estimating a function $f(X) = \bar{y}$, where $\bar{y}$ is the predicted target value that is being compared to the real target value $y$. Thus, the optimization problem is generally defined as minimize$\sum \mathcal{L}(f(X), y)$, where $\mathcal{L}$ is an arbitrary cost function. In this work, we used the mean absolute error cost function which is defined as MAE $= 1/n \sum_{i=1}^{n} |y_i - \bar{y}_i|$. To evaluate the performance of the model, we used data points that had not been used during training and then evaluated them using the cost function with the true values of the target. We used *gradient boosting regression* (GBR) [41], which is basically a decision tree with simple rules that is used for $M$ iterations, where in each iteration a new decision tree is used to predict the previous prediction residual (the GBR outperformed other algorithms and techniques that we tested, such as logistic regression and classical decision trees. The technical details of the GBR algorithm can be found in [41]). We used the scikit-learn [42] Python library implementation of the GBR.

The features we used are shown in Table 1. We used these features to predict cascade size and cascade virality. We used only features from the network $G_0$ and did not use any of the temporal features in order to make the prediction more realistic, as temporal features of a network exhibit proxies for the predicted values, which weakens the applicability of the method. The features described in Table 1 have different effects on the prediction; thus, we performed feature ranking in order to get insights regarding which features are more important during the prediction. Figure 11 shows the feature ranking for predicting cascade size and cascade virality. Figures 11(a) and 11(b) show that the importance of the features is different; for predicting cascade size, the average of *neighbors' degrees* was the most important one, whereas the feature *coreness* was the most importance one for predicting cascade virality. In both cases, the features degree and eccentricity were the least important ones in the set of features. Based on that, we used the five best features from each ranked set. Other combinations of the features resulted in lower, but very close, prediction performance.

To perform a meaningful prediction, we combined the values of all features of the subwebsites used into one dataset. Then, we split this dataset into two subsets, with 75% (1002 cascades) and 25% (334 cascades) for training and testing, respectively. We used the MAE as a prediction accuracy measure. As splitting the dataset into training was done in a random manner, we ran the prediction experiment 100 times to get statistical significance regarding the results. Additionally, we compared the results to a baseline predictor that uses naive rules, such as taking the mean, the median, or a constant value for the predicted target. We compared the prediction results to the best baseline we got, which was the mean baseline. The prediction accuracy of *cascade size* in terms of the MAE was 9.9, which is 35% better than the baseline predictor. The prediction results mean that, on average, the predicted cascade size contains ±10 nodes. The prediction accuracy of *cascade virality* in terms of the MAE was 0.194 which is more than 25% better than the baseline predictor. Figure 8 shows the results of the prediction for the 100 performed runs for predicting both cascade size and the cascade virality in (a) and (b), respectively. The figure shows that there is a clear significance in favor of the GBR algorithm over the baseline predictor.

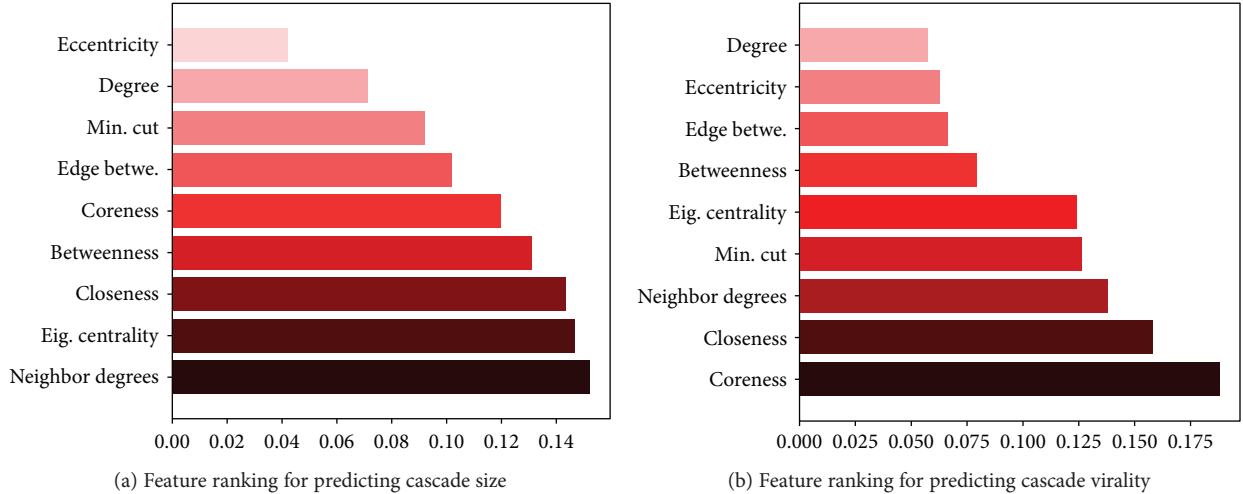

FIGURE 11: The figure shows feature ranking such that $\sum_i w(i) = 1$, where $w(i)$ is the feature rank for the feature $i$. The method used for generating the importance is random forests where the importance of a feature increases whenever a split in the tree using that feature minimizes the prediction error [43].

*Discussion Point 6.* For temporal networks, early network's structure encompasses sufficient information to predict the properties of potential decay cascades.

It was surprising that using only network features from the network $G_0$ provided a satisfactory prediction of cascade's virality and size. These results suggest that the early structure of an evolving network dictates its future. The prediction model described and evaluated in Section 5.2, which used no temporal information at all, indicates that the (in)activity dynamics of social networks is governed by the topological structure of the network itself.

## 6. Closing Thoughts

Although the method used in this work is reliable and the results have been validated, this work is subject to the following limitations. The networks used in this work were aggregated from different types of interactions on Stack Exchange subwebsites. This aggregation used the social interactions among the members of these subwebsites, and we assumed that the resulting network is a community. In order to make sure that the networks we used represent real temporal interaction among the users, we used different time frames to take a snapshot for each subwebsite. The reason for this is that each subwebsite has a different timespan; for example, the alive subwebsites are still active, unlike the decayed subwebsites, which have a significantly shorter lifespan. We believe that our design decisions for selecting the time frames have no significant effect on the results and the conclusions. Also, the results and conclusions in this work are valid for the Stack Exchange subwebsites and similar platforms. We did not check other types of social networks or aimed at generalizing the results to any type of social network. Nor did we provide a model (other than data fitting using the machine learning regression model we described in Section 5.2) for better understanding the decay of online social communities. Such a model might help to eventually control and prevent such decay. This gap remains open and requires future work.

## Data Availability

The dataset and the code used for this work are available upon request.

## Disclosure

This work is part of the PhD ongoing research of Mohammed Abufouda supervised by Prof. Dr. Katharina A. Zweig.

## Conflicts of Interest

The author is employed as research fellow at the computer science department in the University of Kaiserslautern, Germany.

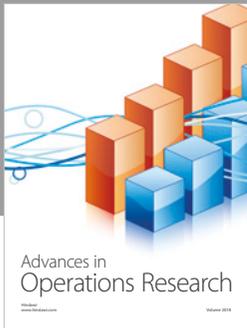
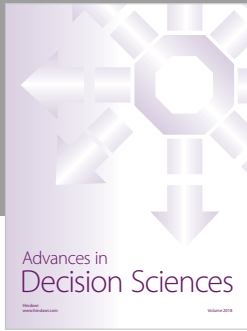
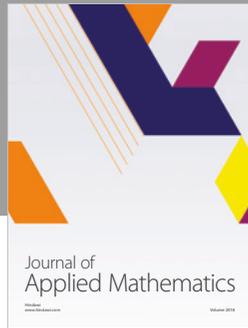
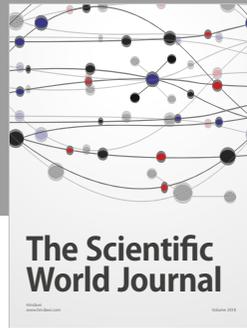
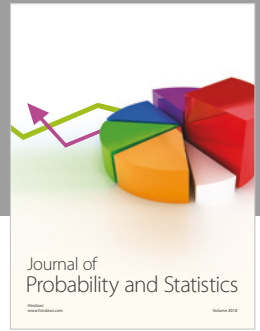
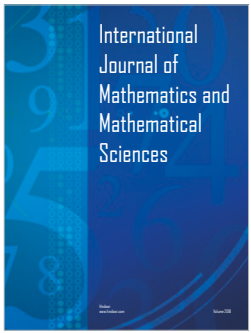
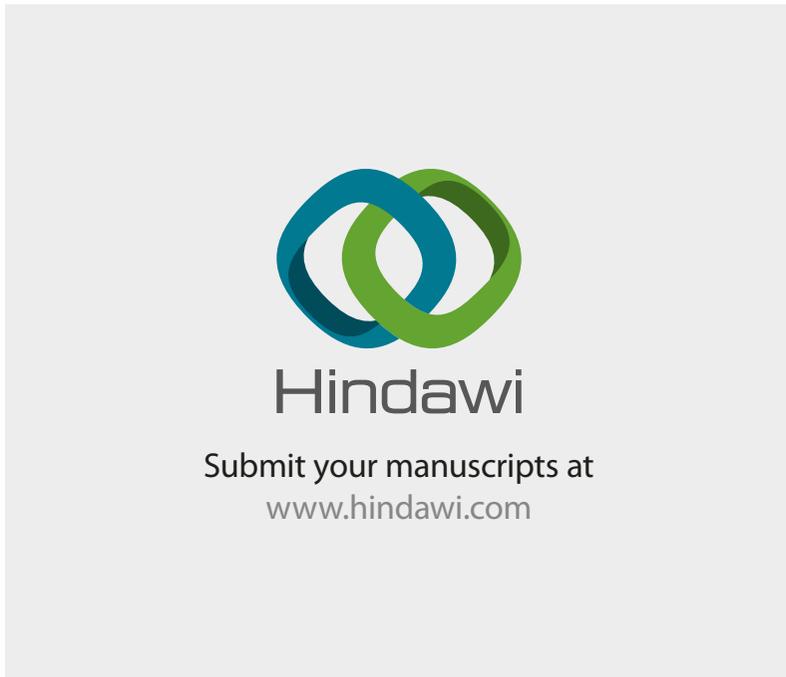
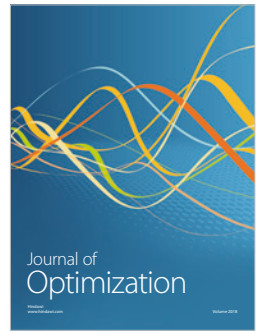
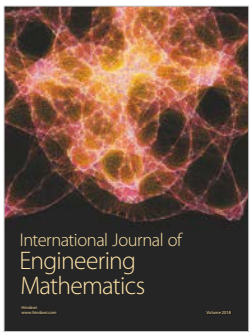
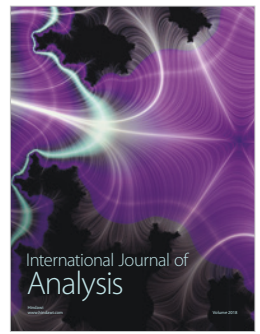
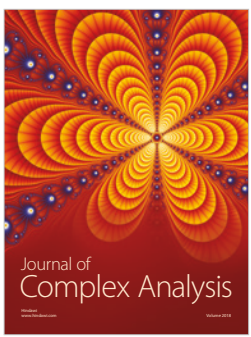
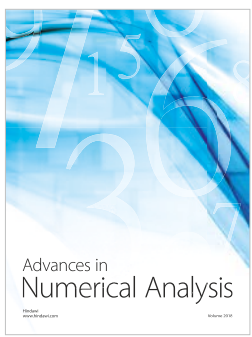
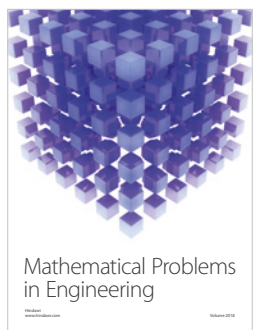
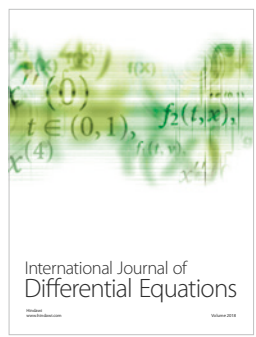
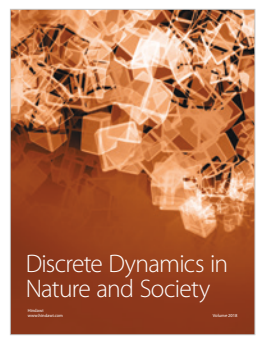
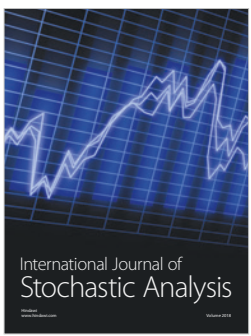
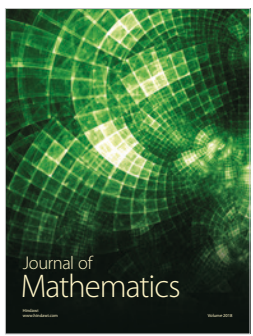
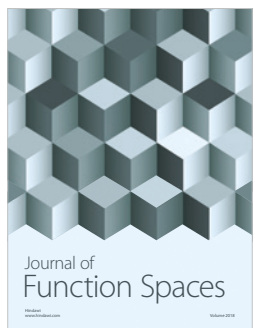
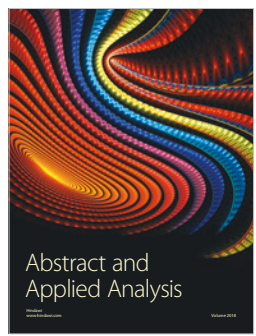
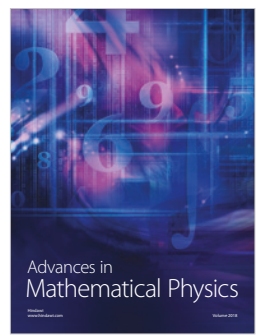

Submit your manuscripts at
www.hindawi.com